\documentclass[aps,prb,twocolumn]{revtex4}   

\usepackage{amsmath}    
\usepackage{graphicx}   

\begin{document}

\title{Simple simulation for electron energy levels\\ in geometrical potential wells}

\author{Teparksorn Pengpan}
 \affiliation{Prince of Songkla University, Department of Physics, PO Box 3, Hat Yai,
Thailand 90112}
 \email{teparksorn.p@psu.ac.th} 
\date{\today}

\begin{abstract}
An octopus program is demonstrated to generate electron energy levels in three-dimensional geometrical potential wells. The wells are modeled to have shapes similar to cone, pyramid and truncated-pyramid. To simulate the electron energy levels in quantum mechanical scheme like the ones in parabolic band approximation scheme, the program is run initially to find a suitable electron mass fraction that can produce ground-state energies in the wells as close to those in quantum dots as possible and further to simulate excited-state energies. The programs also produce wavefunctions for exploring and determining their degeneracies and  vibrational normal modes.
\end{abstract}

\maketitle

\section{Introduction}
Symmetry plays an important role in quantum mechanics.\cite{Shankar,Sakurai} It helps predict and classify wavefunctions a quantum system can have in each energy level. For three-dimensional rotational invariant potentials such as in a harmonic oscillator and a hydrogen atom, their Schr\"{o}dinger's equations,
\begin{equation}
\label{Schrodinger}
-\frac{\hbar ^2}{2m} \vec{\nabla}^2\Psi+V(\vec{r})\Psi =E\Psi,
\end{equation}
can be solved analytically and there exist explicit formulas which relate  a number of wavefunctions or degeneracy to energy level $n$. For other symmetric, geometrical potentials, analytical method may not be able to find  their solutions. Except by guessing from symmetry of the potentials, one hardly knows their approximate solutions. However, with the inclusion of computational tools, their numerical solutions including their degeneracies can be seen explicitly.

As it is realized that simulation is one of the paths toward scientific truth.\cite{Landau,Peskin} In this note, programs in octopus code\cite{Castro,Marques} are demonstrated to generate electron energy levels in three-dimensional geometric potential wells. Although octopus package is designed to solve many-body systems based on time-dependent density functional theory, it is able to solve single-particle systems based on the Schr\"{o}dinger's equation and to produce output files for potentials and wavefunctions, which can be viewed by visualizers such as Gnuplot, OpenDX, etc. The potential wells are modeled to have shapes similar to cone, pyramid and truncated-pyramid.\cite{Li,Hwang,Wang,Voss} To simulate the electron energy levels in quantum mechanical scheme like the ones in parabolic band approximation scheme, the programs are run initially to find ground-state energies in the wells as close to those in quantum dots as possible and further to simulate excited-state energies. The programs also produce wavefunctions, which are visualized by OpenDX, for exploring and determining their degeneracies and vibrational normal modes.

\section{Symmetric potentials}
Before doing quantum mechanical simulation for three-dimensional potential wells, it would be nice to test the precision of octopus with well-known symmetric potentials. Let one try first with one-dimensional harmonic oscillator potential,
\begin{equation}
V(x) = \frac 12 x^2,
\end{equation} 
and one-dimensional symmetrical linear potential,
\begin{equation}
\label{eq:sym-linear}
V(x) = |x|.
\end{equation}
A bonus from the symmetric linear potential is that one is able to derive solutions for the linear one,
\begin{equation}
\label{eq:linear}
V(x) =
\begin{cases}
x & \text{$x > 0$} \\
\infty & \text{$x \leq 0,$}
\end{cases}
\end{equation}
by symmetry consideration. Although the linear potential is simple, one needs to look for zeros of Airy function to get energy eigenvalues.\cite{Sakurai,Langhoff}
\subsection{The octopus input}
For demonstration in this note, the octopus source code version 3.0.1\cite{octopus} is downloaded and complied by gfortran on 64-bit Xubuntu 8.04 LTS running on Intel E7200 with 4GB RAM.  The input file for a particle called ``Q-ball" in the harmonic oscillator potential (\ref{eq:sym-linear}) is shown below.
\begin{verbatim}
 (1) %CalculationMode
 (2)    gs | unocc
 (3) %
 (4) NumberUnoccStates = 9
 (5) TheoryLevel = independent_particles
 (6) EigenSolverMaxIter = 5000
 (7) Dimensions = 1 
 (8) radius = 10
 (9) spacing = 0.1
(10) % Species 
(11) "Q-ball"|1|user_defined|1|"0.5*x^2" 
(12) %
(13) % Coordinates
(14) "Q-ball" | 0 | 0 | 0 
(15) %
(16) Output = wfs + potential
(17) OutputHow = axis_x + gnuplot 
(18) OutputWfsNumber = "1-10"
\end{verbatim}
Line numbers are associated with the following references. 
\begin{itemize}
\item Line 1--3  form a block of calculation mode. The program will run initially for a ground state and further for the unoccupied states.
\item Line 5--6 declare type of theory and a maximum iteration number in solving for each eigenvalue.
\item Line 7--9 declare a dimension, a half-length of computational box and a spacing number or mesh size.
\item Line 10--12 form a block defining a particle called ``Q-ball" having one electron mass and charge in a user-defined potential $0.5x^2$.
\item Line 13--15 form a block defining Q-ball's coordinates.
\item Line 16--18 declare output options.
\end{itemize}
For the symmetric linear potential, just replace the harmonic oscillator potential with abs(x). After discretizing the Hamiltonian over mesh points, eigenvalues are solved by using conjugate gradient method. Tolerance of each eigenvalue is of order $10^{-6}$ by octopus' default setting.
\subsection{The octopus output}
\begin{figure}[h]
\begin{center}
\includegraphics[width=1.6in]{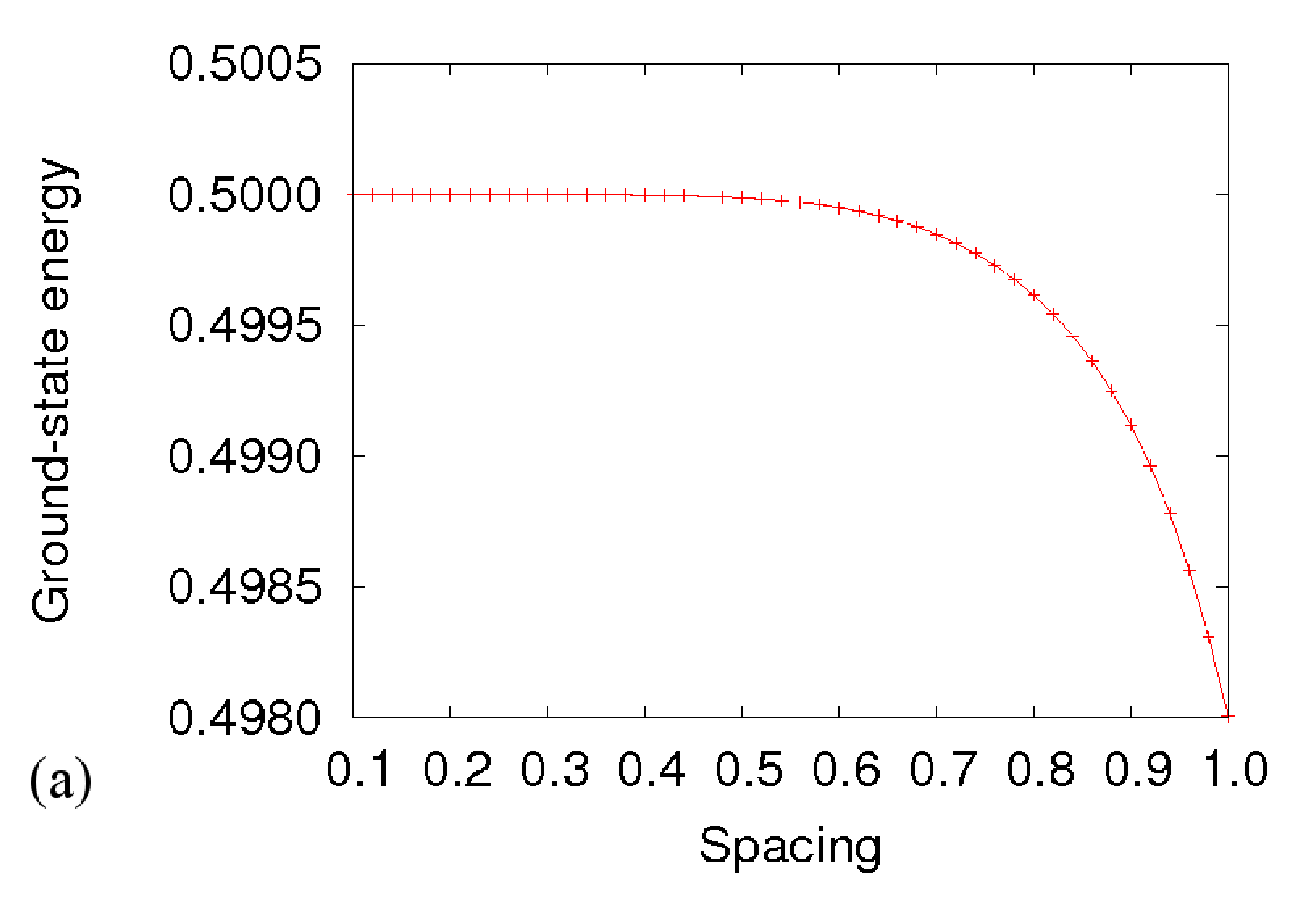}
\includegraphics[width=1.6in]{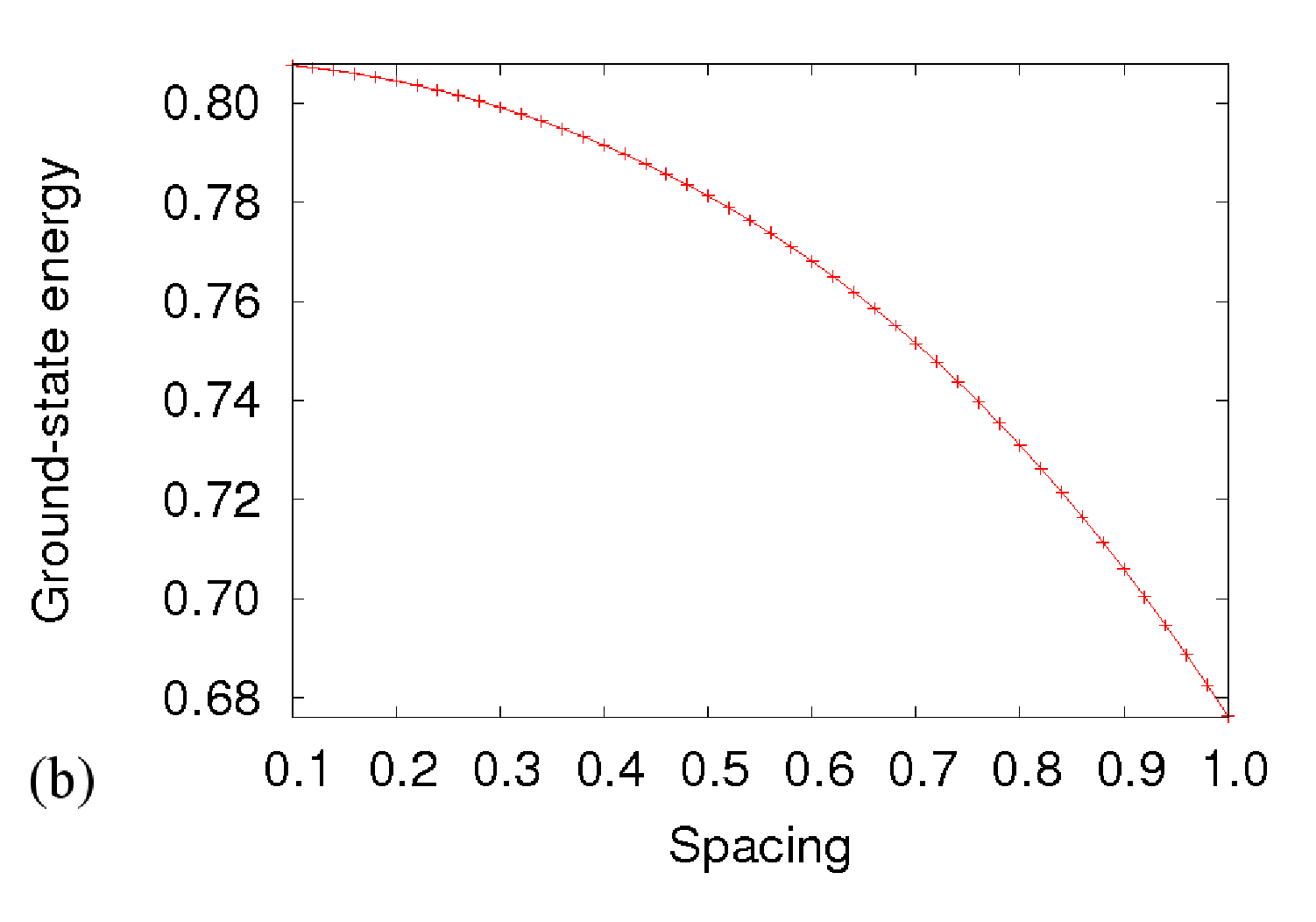}
\caption{\label{fig:1d-energy-spacing}Ground-state energy vs spacing of (a) harmonic oscillator potential and (b) symmetric linear potential.}
\end{center}
\end{figure}

Since octopus is based on real-space calculation, a spacing number between two nearest-neighbored  mesh points is one of the key parameters. So, before running the aforementioned input, it needs to check whether the chosen spacing is able to produce precise energy eigenvalues. The ground-state energies  of both the harmonic oscillator and symmetric linear potentials varying with spacings are shown in Fig.~\ref{fig:1d-energy-spacing}. Spacings starting from 0.2 give ground-state energies of the harmonic oscillator gradually falling off from their exact value of 0.5 ($\hbar = \omega = 1$), while those of the symmetric linear potential rapidly falling off from their upper bound value. It is noted that  for the harmonic oscillator the spacing 0.1 produces the exact energy eigenvalues up to $n=8$ and slightly off the exact value when $n >8 $. Energy eigenvalues of both the harmonic oscillator and the symmetric linear potentials from octopus are shown in Table~\ref{tab:linear-energy-level}. 
\begin{table}[h]
\begin{center}
\begin{tabular}{|c|c|c|c|c|}
\hline 
level & harmonic &symmetric &  linear & zeros of \\
 $n$ & oscillator & linear &  & Airy function\cite{AS} \\
\hline
1  & 0.500000 & 0.807584 & 1.855759 & 2.338107 \\   
2  & 1.500000 & 1.855759 & 3.244609 & 4.087949 \\    
3  & 2.500000 & 2.577771 & 4.381673 & 5.520560 \\    
4  & 3.500000 & 3.244609 & 5.386615 & 6.786708 \\    
5  & 4.500000 & 3.825496 & 6.305265 & 7.944134 \\   
6  & 5.500000 & 4.381673 &    &   \\   
7  & 6.500000 & 4.891648 &    &   \\    
8  & 7.500000 & 5.386615 &    &   \\   
9  & 8.499999 & 5.851157 &    &   \\    
10 & 9.499999 &  6.305265 &    &   \\       
\hline
\end{tabular}
\caption{\label{tab:linear-energy-level}Energy eigenvalues of harmonic oscillator potential, symmetrical linear potential, linear potential and zeros of Airy function. When the energy eigenvalues of the linear potential are multiplied by $\sqrt[3]{2}$, the resulting values are approximately equal to zeros of the Airy function.\cite{AS}}
\end{center}
\end{table}
Energy eigenvalues of the linear potential are derived from those of the symmetric one by imposing a boundary condition that wavefunctions must be zero at $x=0$. Hence, only energy eigenvalues of the symmetrical linear potential at even-number level are those of the linear one, shown in the third column of Table~\ref{tab:linear-energy-level}.

Now, the input can be extended from one-dimensional to three-dimensional by setting `Dimensions = 3' and `radius = 8' and replacing the potential with 
\begin{verbatim}
0.5*(x^2+y^2+z^2)
\end{verbatim}
Variation of the ground-state energies with spacings of three-dimensional harmonic oscillator has the same pattern as that of the one-dimensional; the energy eigenvalues gradually fall off its exact value of 1.5 as spacings increase. 
\section{Geometrical potential wells}
\begin{figure}[h]
\begin{center}
\includegraphics[width=3.2in]{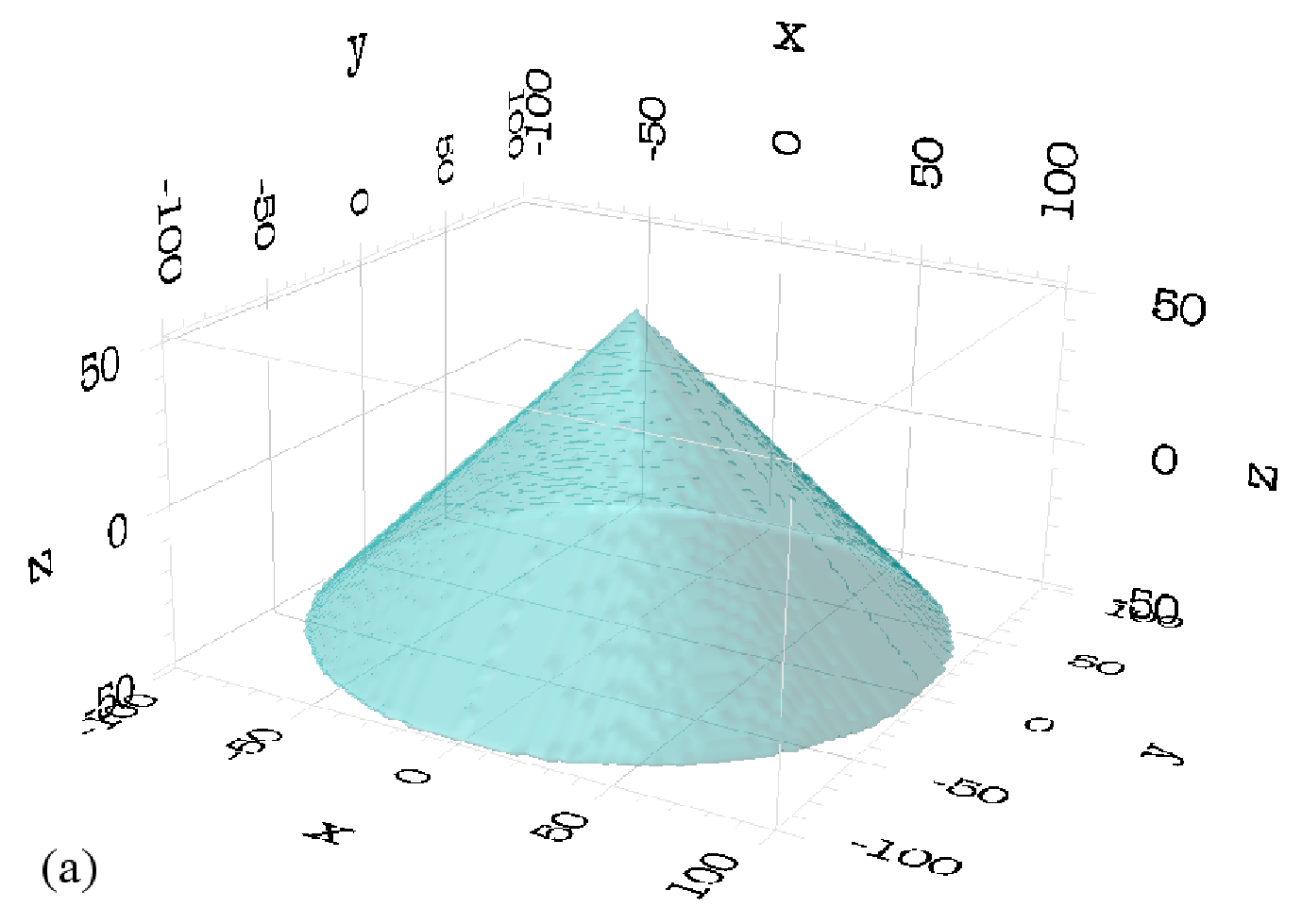} \\
\includegraphics[width=3.2in]{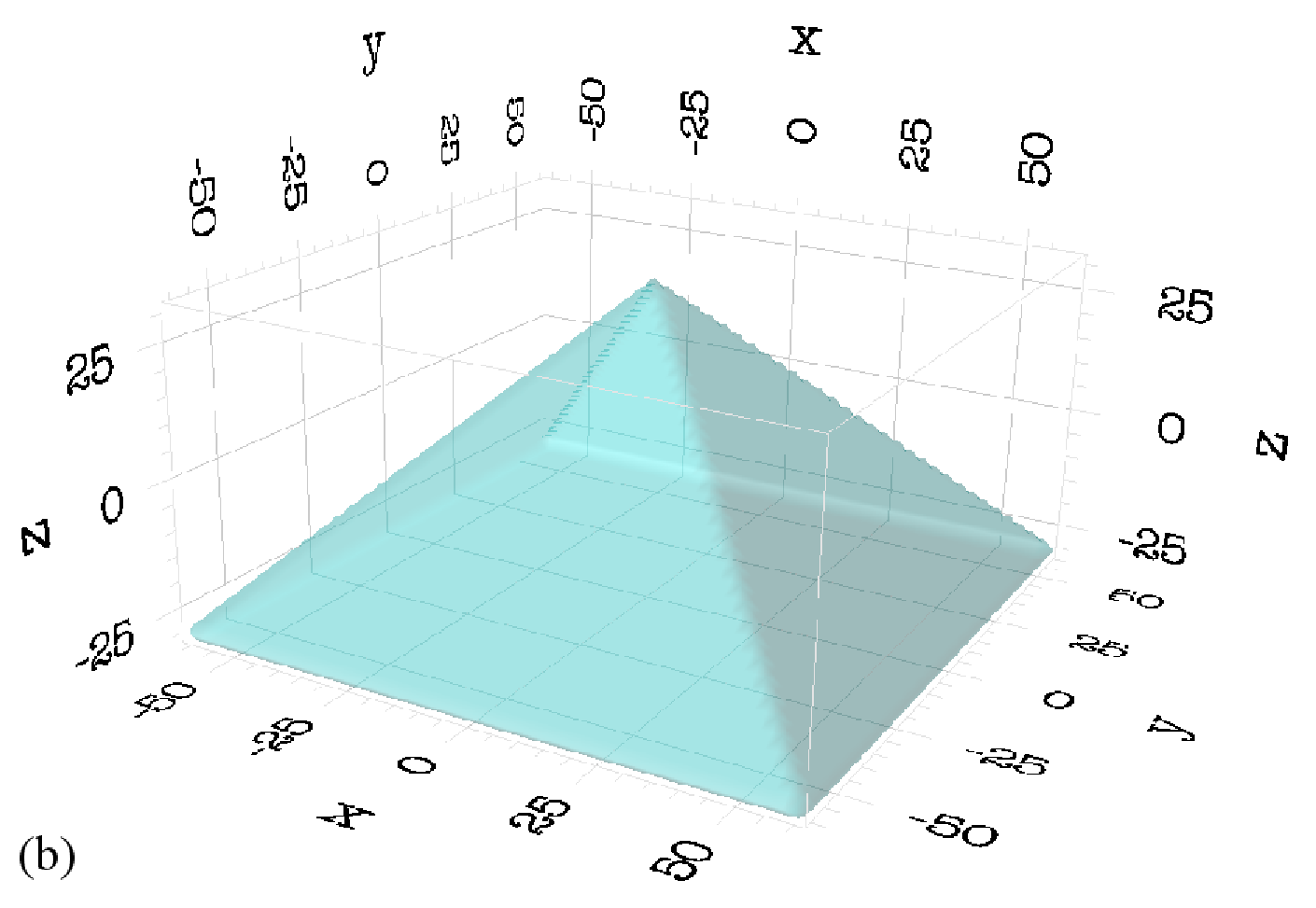} \\
\includegraphics[width=3.2in]{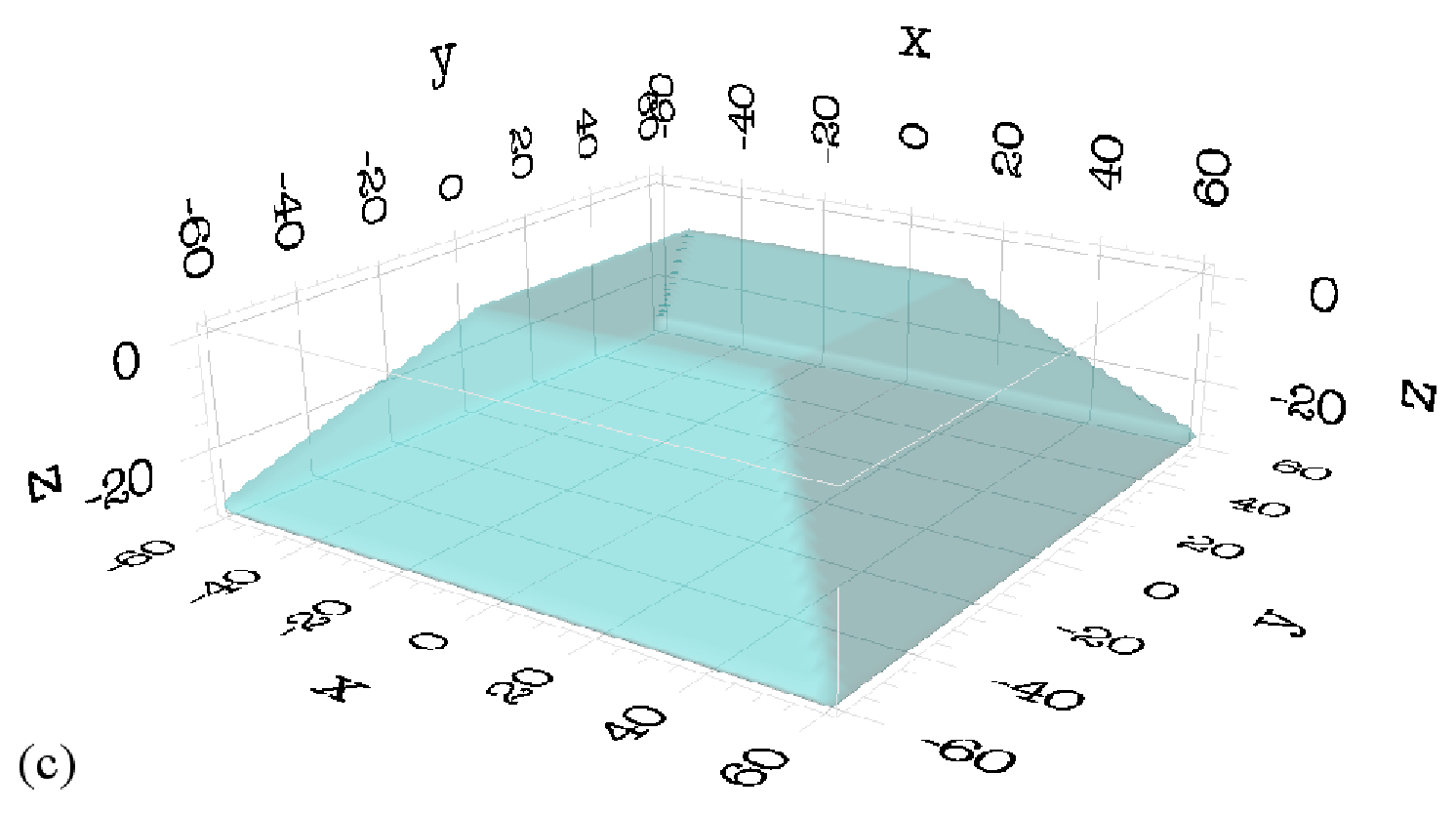}
\caption{\label{fig:geo-potentials}Geometric potential wells in the shape of (a) cone, (b) pyramid and (c) truncated-pyramid. These pictures are produced by octopus and viewed by OpenDX.}
\end{center}
\end{figure}

Next, we will do simulation for geometrical potential wells. Their shapes shown in Fig.~\ref{fig:geo-potentials} are a cone of radius 10 nm and height 10 nm,\cite{Li,Voss} a pryramid  of width 12.4 nm and height 6.2 nm,\cite{Hwang,Voss} and a truncated-pyramid of width 12.4 nm and height 3.1 nm.\cite{Wang} 
\subsection{The octopus input}
An input file for octopus to simulate electron energy levels and wavefunctions in a geometric potential well V(x,y,z) with depth offset of $-0.77$ is shown below. 
\begin{verbatim}
 (1) Units = ev_angstrom		
 (2) CalculationMode = gs
 (3) # CalculationMode = unocc  
 (4) # NumberUnoccStates = 11  	
 (5) TheoryLevel = independent_particles 			
 (6) # EigenSolverMaxIter = 5000	
 (7) Dimensions = 3
 (8) BoxShape = parallelepiped		
 (9) lsize = 150	
(10) spacing = 3.785			
(11) ParticleMass = 0.036 			
(12) % Species 				
(13) "QD"|ParticleMass|user_defined|1|"V(x,y,z)" 
(14) %  
(15) % Coordinates		
(16) "QD" | 0 | 0 | 0  
(17) % Discussions
(18) Output = potential + wfs	
(19) OutputHow = dx  		
(20) OutputWfsNumber = "1-12"  
\end{verbatim}
The `lsize' and `V(v,y,z)' in the input are replaced according to the following potential wells:  
\begin{itemize} 
\item Cone \\
Set `lsize' in the following block form
\begin{verbatim}
%Lsize
140 | 140 | 80
%
\end{verbatim}
and replace `V(x,y,z)' with
\begin{verbatim}
-0.77*step(50-abs(z))
*step(50-sqrt(x^2+y^2)-z)
\end{verbatim}
\item Pyramid \\
Set `lsize' in the following block form
\begin{verbatim}
%Lsize
124 | 124 | 60
%
\end{verbatim}
and replace `V(x,y,z)' with
\begin{verbatim}
-0.77*step(62-abs(x))*step(62-abs(y))
*step(31-abs(z))*step(31-z-abs(x))
*step(31-z-abs(y))
\end{verbatim}
\item Truncated-pyramid \\
Replace the pyramid potential with
\begin{verbatim}
-0.77*step(62-abs(x))*step(62-abs(y))
*step(31+z)*step(31-z-abs(x))
*step(-z)*step(31-z-abs(y))
\end{verbatim}
\end{itemize}
Due to regular grid discretization in all directions, there is only one spacing parameter. By using octopus' default settings, eigenvalues are solved by conjugate gradient method and their tolerance is of order $10^{-6}$.

\subsection{The octopus output}
\begin{figure}[h]
\begin{center}
\includegraphics[width=3.2in]{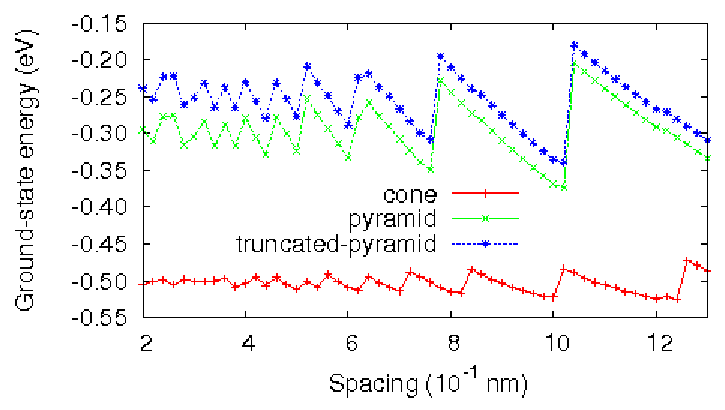}
\caption{\label{fig:geo-energy-spacing}Ground-state energy vs spacing.}
\end{center}
\end{figure}
\begin{figure}[h]
\begin{center}
\includegraphics[width=3.2in]{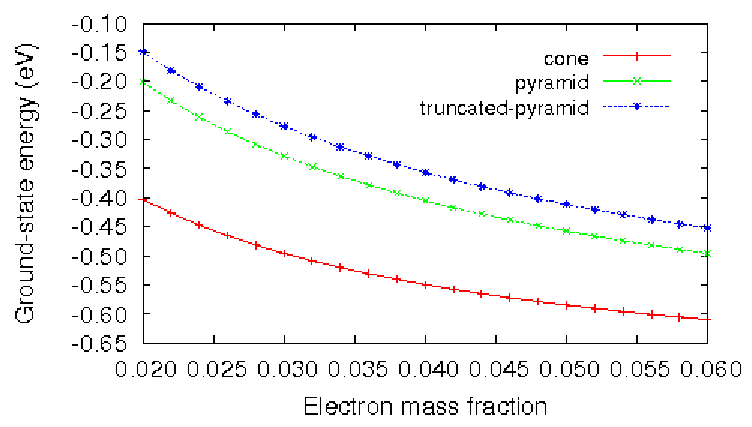}
\caption{\label{fig:geo-energy-mass}Ground-state energy vs electron mass fraction.}
\end{center}
\end{figure}
\begin{figure}[h]
\begin{center}
\includegraphics[width=3.2in]{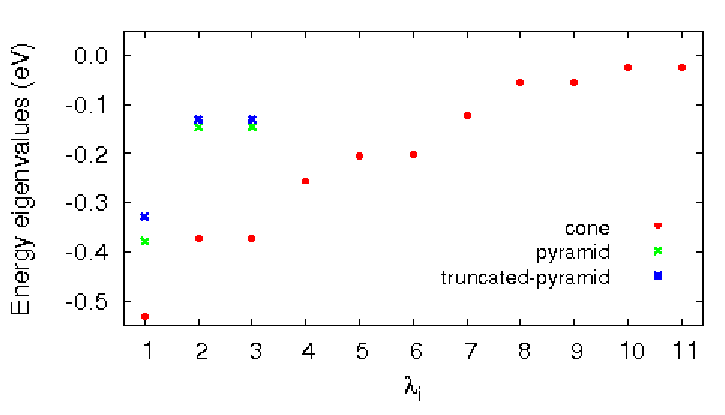}
\caption{\label{fig:geo-energy-level}Eigenvalue distributions in the cone potential, the pyramid potential and the truncated-pyramid potential.}
\end{center}
\end{figure}
The input files are run in the following steps:
\begin{enumerate}
\item Explore how the ground-state energies vary with spacings. To run the input automatically, use a shell script below.
\begin{verbatim}
#!/bin/bash
list="2 3 4 5 6 7 8 
      9 10 11 12 13"
export OCT_PARSE_ENV=1
for OCT_spacing in $list
do
 export OCT_spacing
 octopus >& out-$OCT_spacing
 energy=`grep Total static/info \
    | head -1 | cut -d "=" -f 2`
 echo $OCT_spacing $energy
 rm -rf restart
done
\end{verbatim}
Unexpectedly, when spacings increase, ground state energies in the wells vary in sawtooth patterns with increasing amplitudes and period widths as shown in Fig.~\ref{fig:geo-energy-spacing}. It seems that for each potential well there are many spacing parameters that can produce the same ground-state energy. To simulate the energy eigenvalues as those in Refs.~\onlinecite{Hwang} and \onlinecite{Wang}, the spacing parameter of 0.3875 nm is chosen. Due to time limit in class for demonstration, higher spacing parameters such as 0.775 nm is recommended.
\item Find a suitable electron mass fraction that can produce ground-state energies in the geometrical potential wells as close to those in the quantum dots as possible.\cite{Hwang,Wang,Voss}  To run them automatically by using the previous script, replace numbers in the `list' with the ones as shown on the horizontal axis of Fig.~\ref{fig:geo-energy-mass} and `spacing' with `ParticleMass' everywhere. As shown in Fig.~\ref{fig:geo-energy-mass}, the ground-state energies smoothly decrease as mass fractions increase. It is estimated that the suitable mass fraction is 0.036.
\item Explore the excited-state energies and wavefunctions by choosing the `unocc' mode. The distributions of the energy eigenvalues in the geometrical potential wells are shown in Fig.~\ref{fig:geo-energy-level}. The wavefunctions are exported as DX-formatted two-dimensional cross sections for exploring their normal modes:
\begin{itemize}
\item cone\\
\begin{figure}[h]
\begin{center}
\includegraphics[width=1.6in]{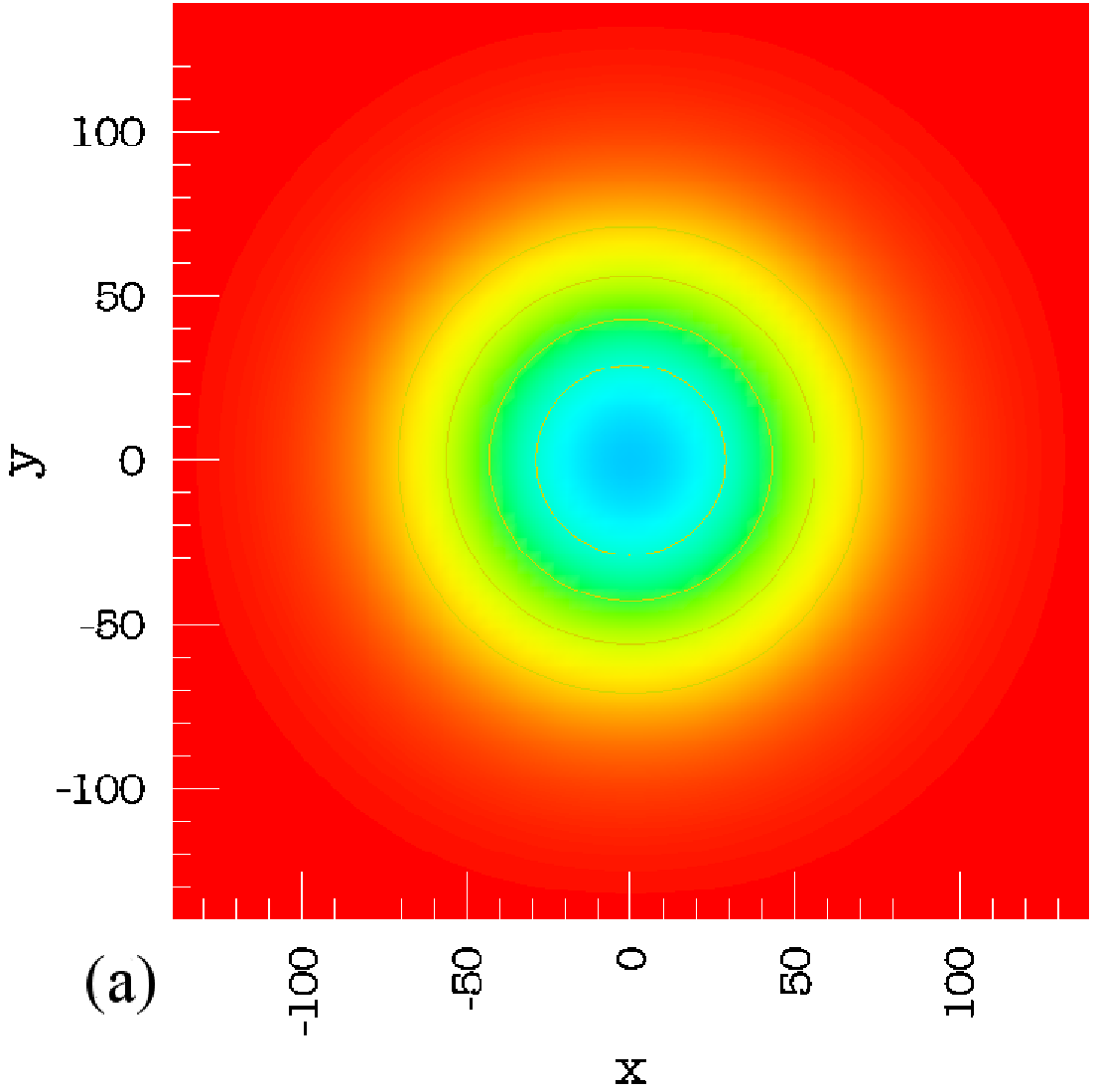}
\includegraphics[width=1.6in]{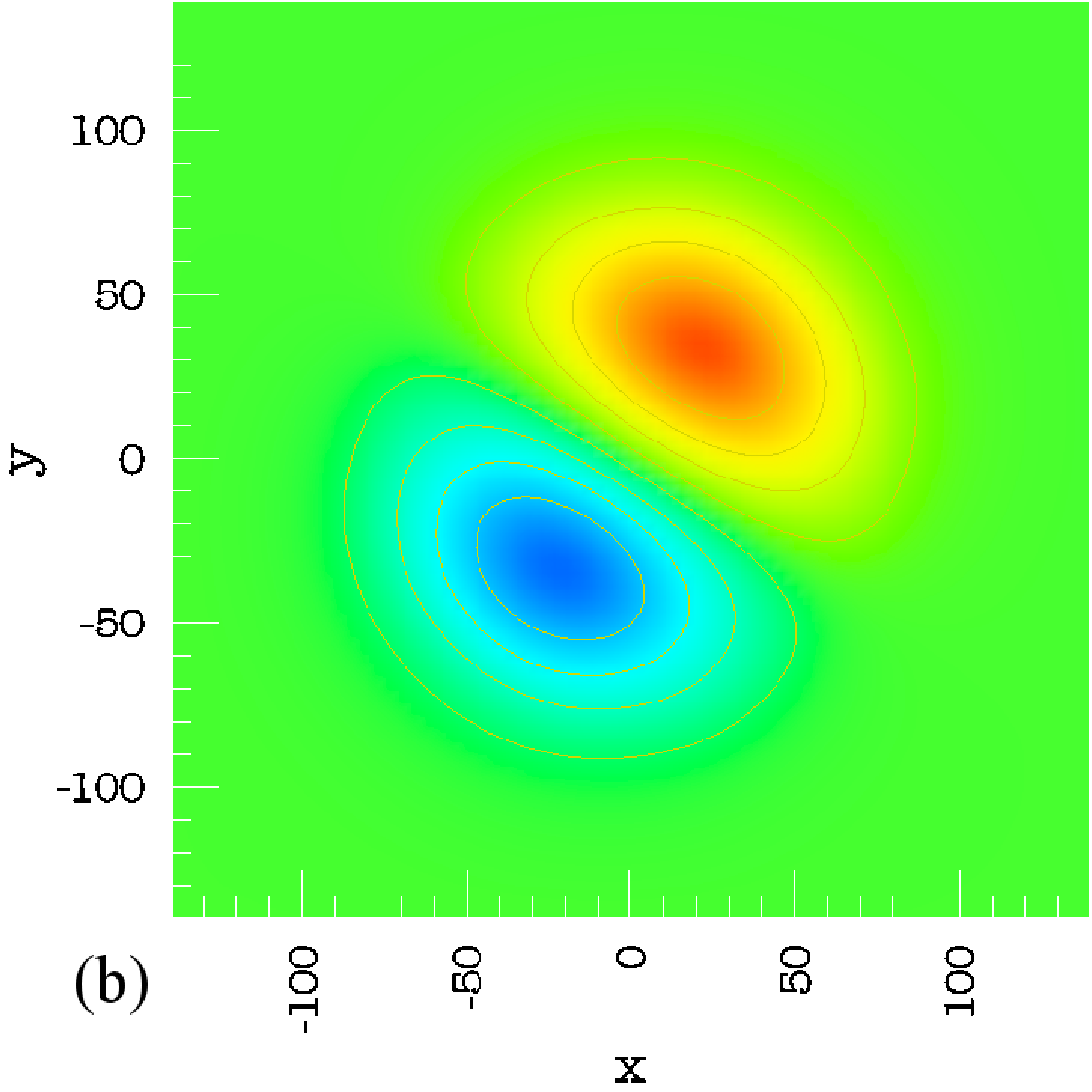}
\includegraphics[width=1.6in]{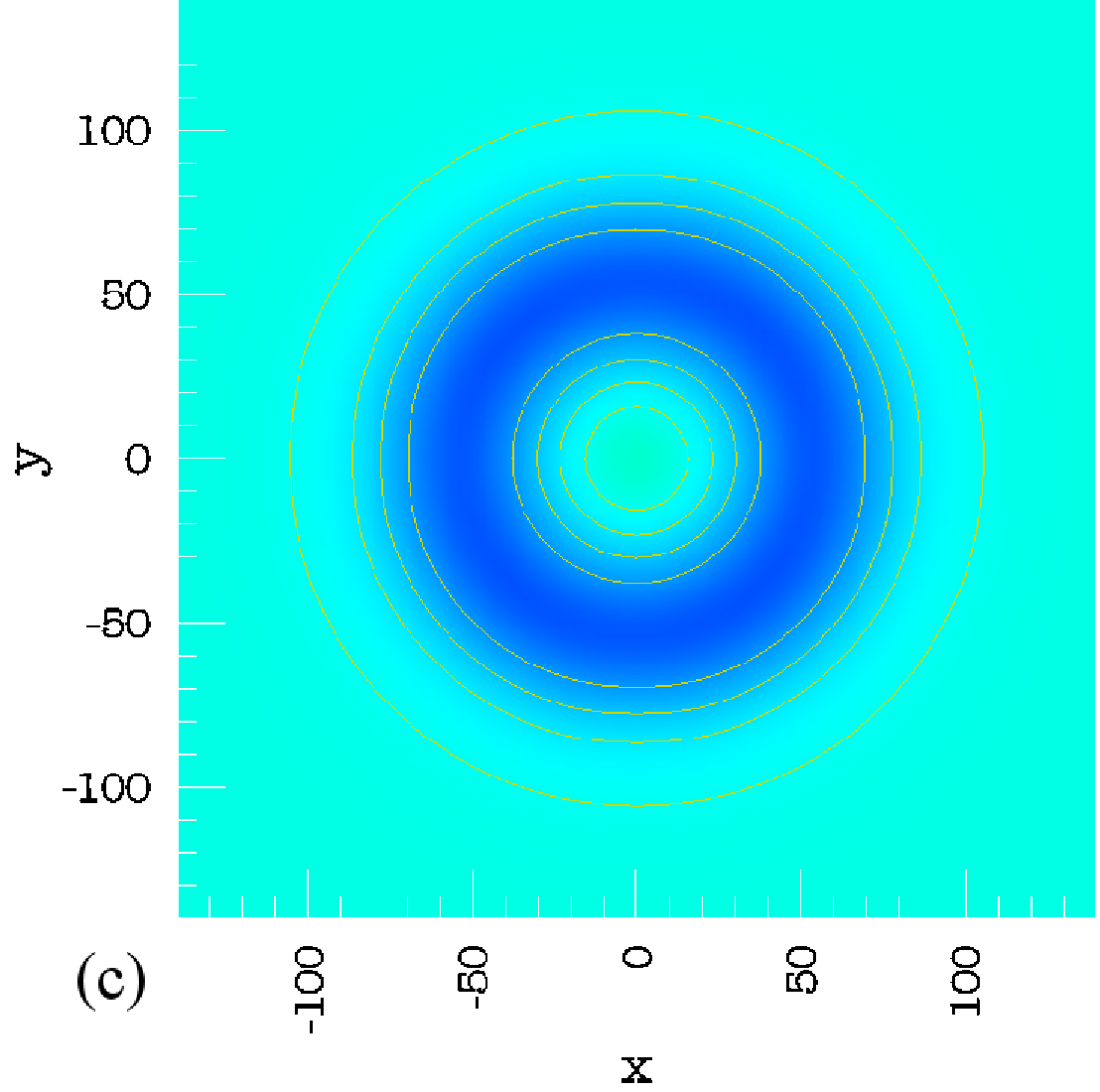}
\includegraphics[width=1.6in]{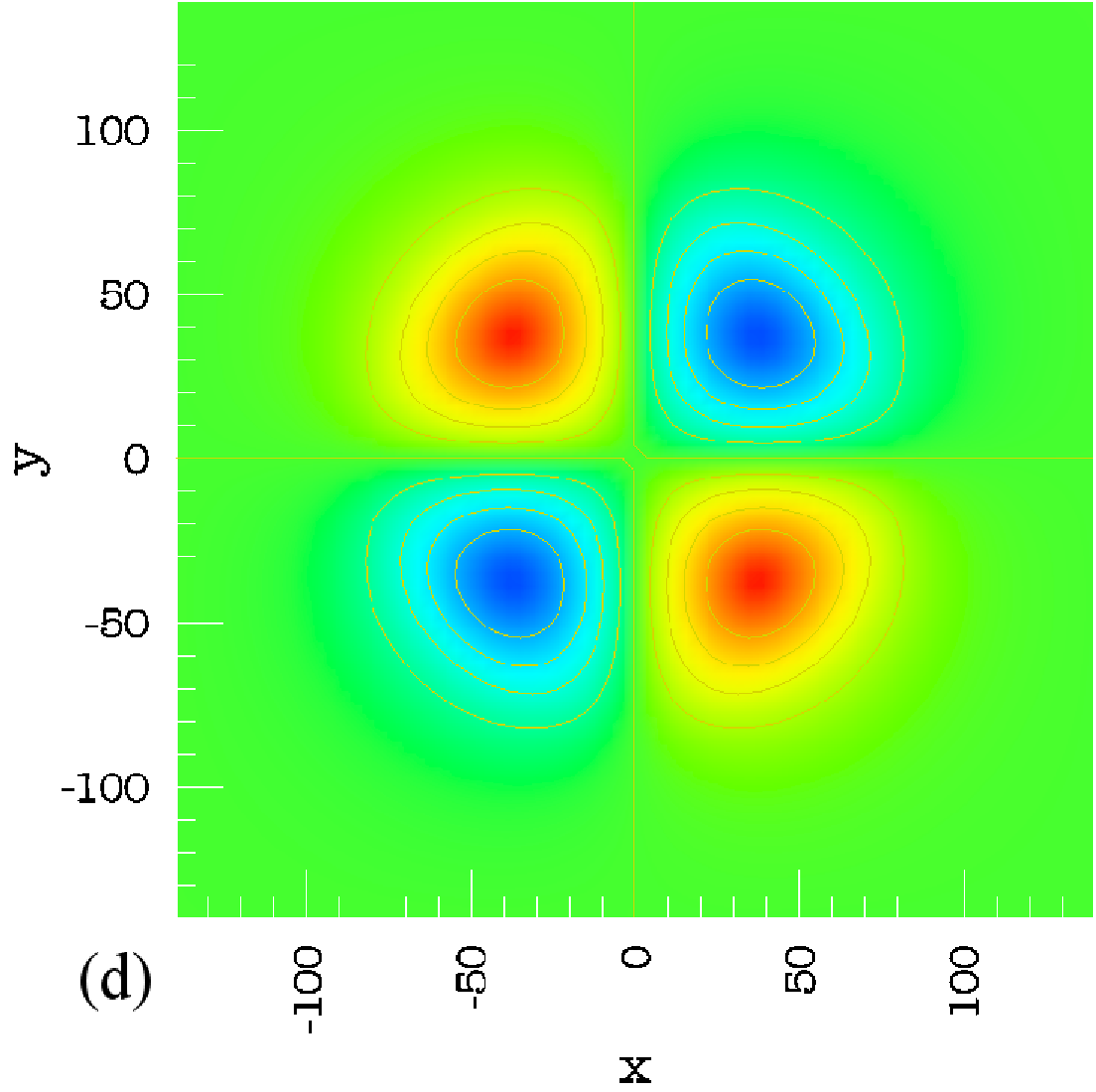}
\includegraphics[width=1.6in]{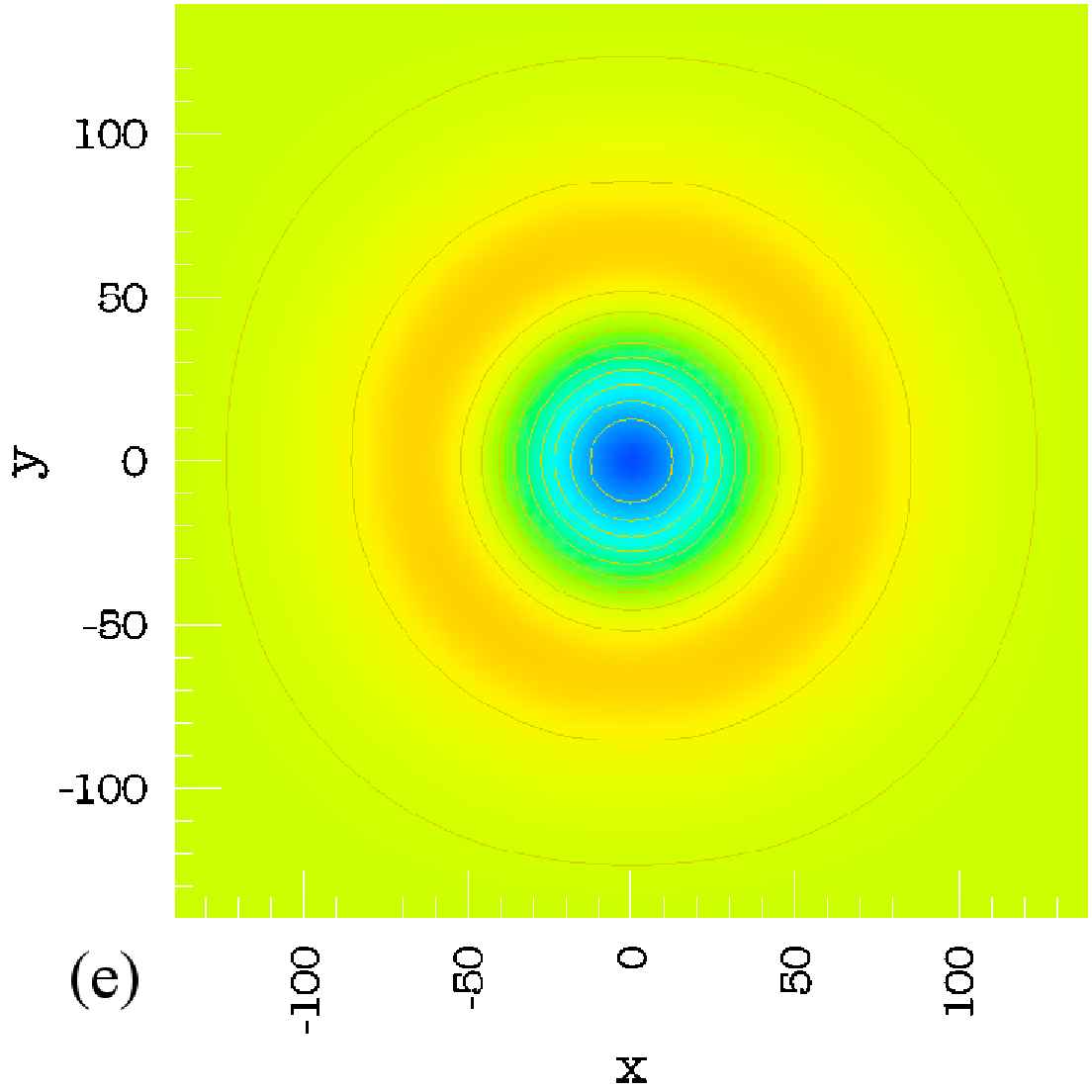}
\includegraphics[width=1.6in]{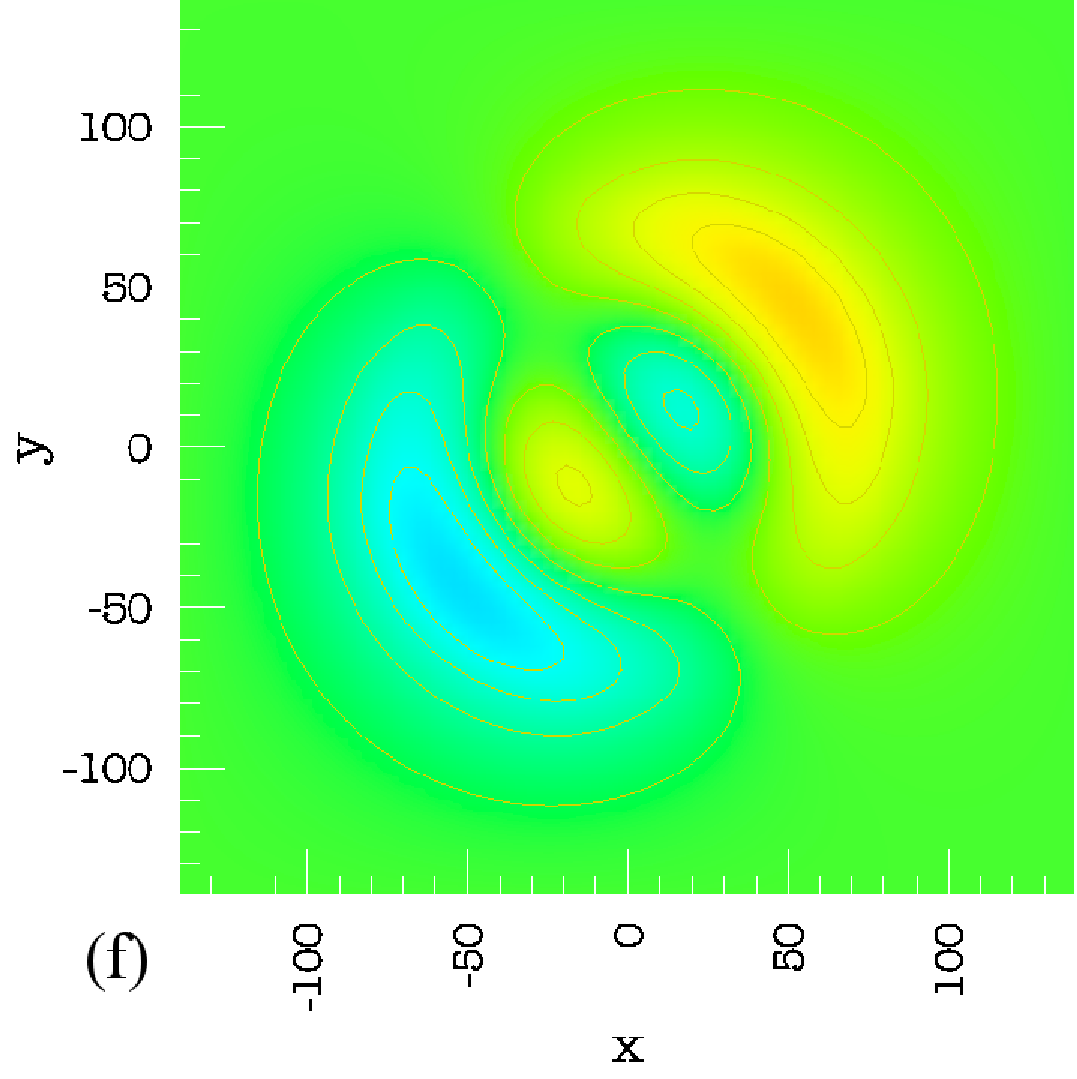}
\includegraphics[width=1.6in]{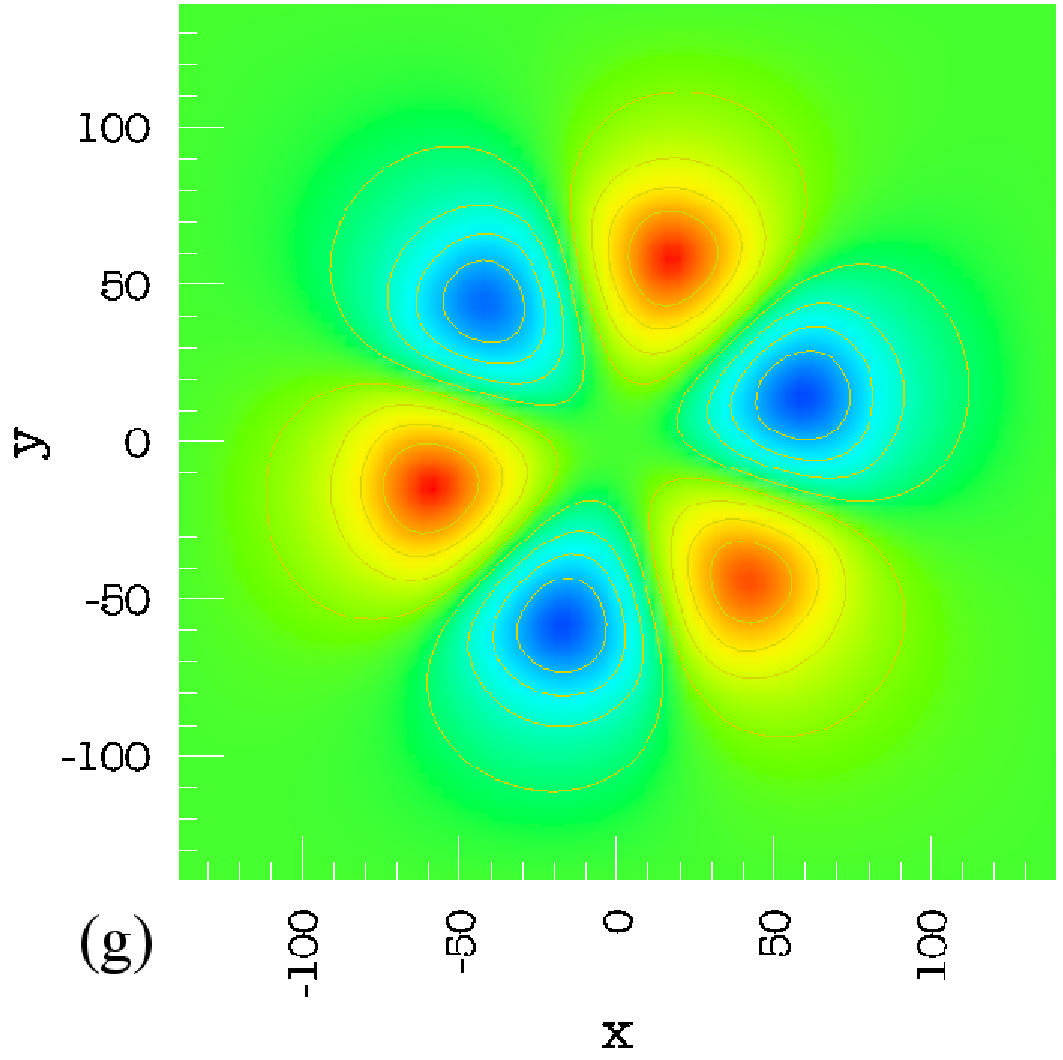}
\caption{\label{fig:cone-wfs}Cross sections of (a) $\Psi_{\lambda_1}$, (b) $\Psi_{\lambda_2}$, (c) $\Psi_{\lambda_4}$, (d) $\Psi_{\lambda_5}$, (e) $\Psi_{\lambda_7}$, (f) $\Psi_{\lambda_8}$ and (g) $\Psi_{\lambda_{10}}$ in the cone potential at $\text{z} = -31$.}
\end{center}
\end{figure}
Cross sections of wavefunctions in accordance with energy eigenvalues $\lambda_1$, $\lambda_2$, $\lambda_4$, $\lambda_5$, $\lambda_7$, $\lambda_8$ and  $\lambda_{10}$ are shown in Fig.~\ref{fig:cone-wfs}. The wavefunctions of $\lambda_3$ and $\lambda_9$ are obtained from those of $\lambda_2$ and $\lambda_8$ by rotating around the $\hat{z}$-axis $90^o$ clockwise and counter-clockwise, respectively, and the ones of $\lambda_6$ and $\lambda_{11}$ from those of $\lambda_5$ and $\lambda_{10}$ by rotating around the $\hat{z}$-axis $45^o$ clockwise and counter-clockwise, respectively. It appears that all two-dimensional cross sections of wavefunctions in the cone look similar to the vibrational normal modes of a circular membrane\cite{Croxton}
\begin{equation}
\label{eq:circular-membrane}
 \Psi_{(n,m)} \sim J_n(k_m\rho)
\begin{cases}
\sin (n\phi) \\
\cos(n\phi)
\end{cases}
\end{equation}
where subscript $(n,m)$ denotes a vibrational normal mode of the wavefunction with harmonic order $m$th of the $n$th-order Bessel function of the first kind $J_n(k_m\rho)$. The normal mode $(n,m)$ of the circular membrane is in general doubly degenerate except $(0,m)$ which is non-degenerate. Hence, the normal mode of the wavefunction in the cone $\Psi_{\lambda_i}$ matches exactly one--to--one with that in the circular membrane $\Psi_{(n,m)}$. It yields the following equivalence in normal modes of the wavefunctions: 
\begin{eqnarray}
&&\Psi_{\lambda_1}  \sim  \Psi_{(0,1)},~~ \Psi_{\lambda_{2,3}} \sim \Psi_{(1,1)},~~ \Psi_{\lambda_4} \sim \Psi_{(0,2)}, \nonumber\\
&&\Psi_{\lambda_{5,6}}  \sim \Psi_{(2,1)},~~\Psi_{\lambda_7} \sim \Psi_{(0,3)},~~\Psi_{\lambda_{8,9}} \sim \Psi_{(1,2)}, \nonumber\\
&& \Psi_{\lambda_{10,11}}  \sim  \Psi_{(3,1)}. 
\end{eqnarray}
\item pyramid and truncated-pyramid \\
Cross sections of wavefunctions in accordance with energy eigenvalues $\lambda_1$ and  $\lambda_2$ in the pyramid and the truncated-pyramid are shown in Fig.~\ref{fig:pyramid-wfs} and \ref{fig:trunc-wfs}, respectively. Both wavefunctions of $\lambda_3$ are obtained from those of $\lambda_2$ by rotating around the $\hat{z}$-axis $90^o$ counter-clockwise and clockwise, respectively. It also appears that all two-dimensional cross sections of wavefunctions in the pyramid and the truncated-pyramid look similar to vibrational normal modes of a square membrane\cite{Croxton}
\begin{figure}[h]
\begin{center}
\includegraphics[width=1.6in]{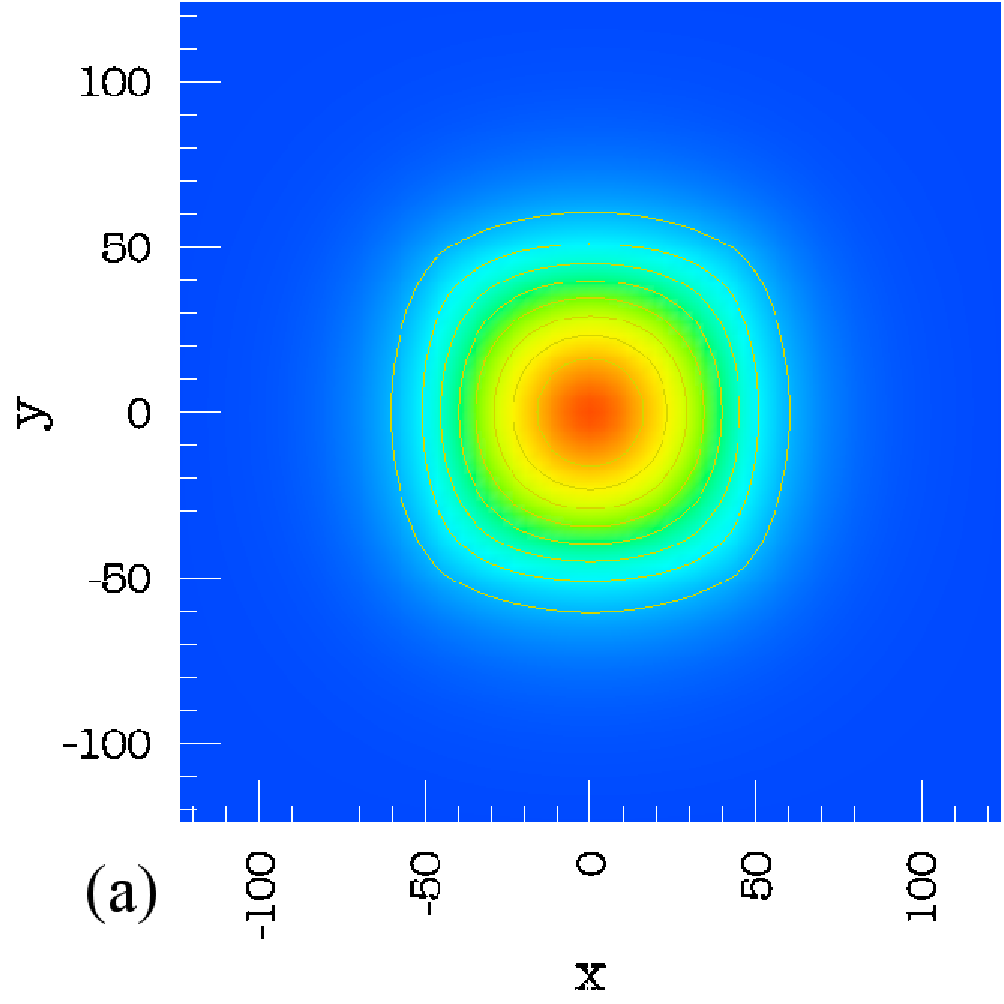}
\includegraphics[width=1.6in]{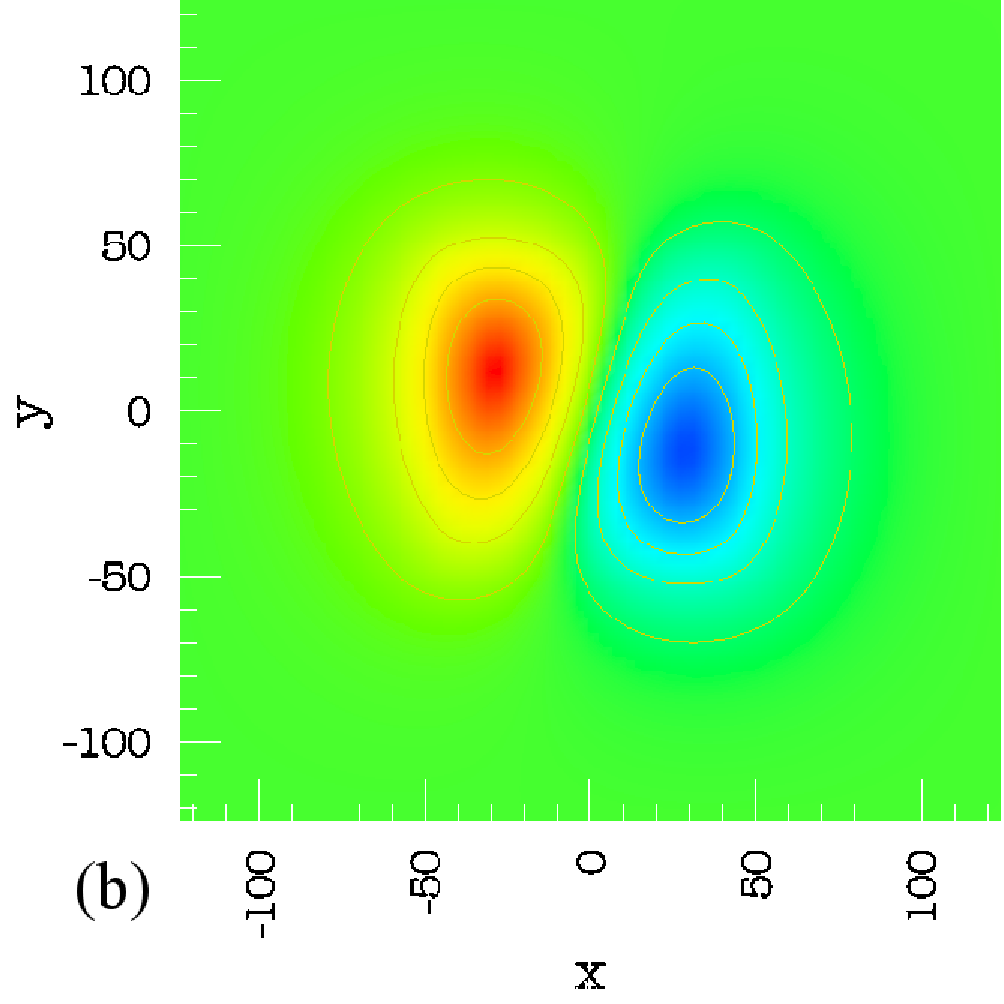}
\caption{\label{fig:pyramid-wfs}Cross sections of (a) $\Psi_{\lambda_1}$ and (b) $\Psi_{\lambda_2}$ in the pyramid potential at $\text{z} = -15.5$.}
\end{center}
\end{figure}
\begin{figure}[h]
\begin{center}
\includegraphics[width=1.6in]{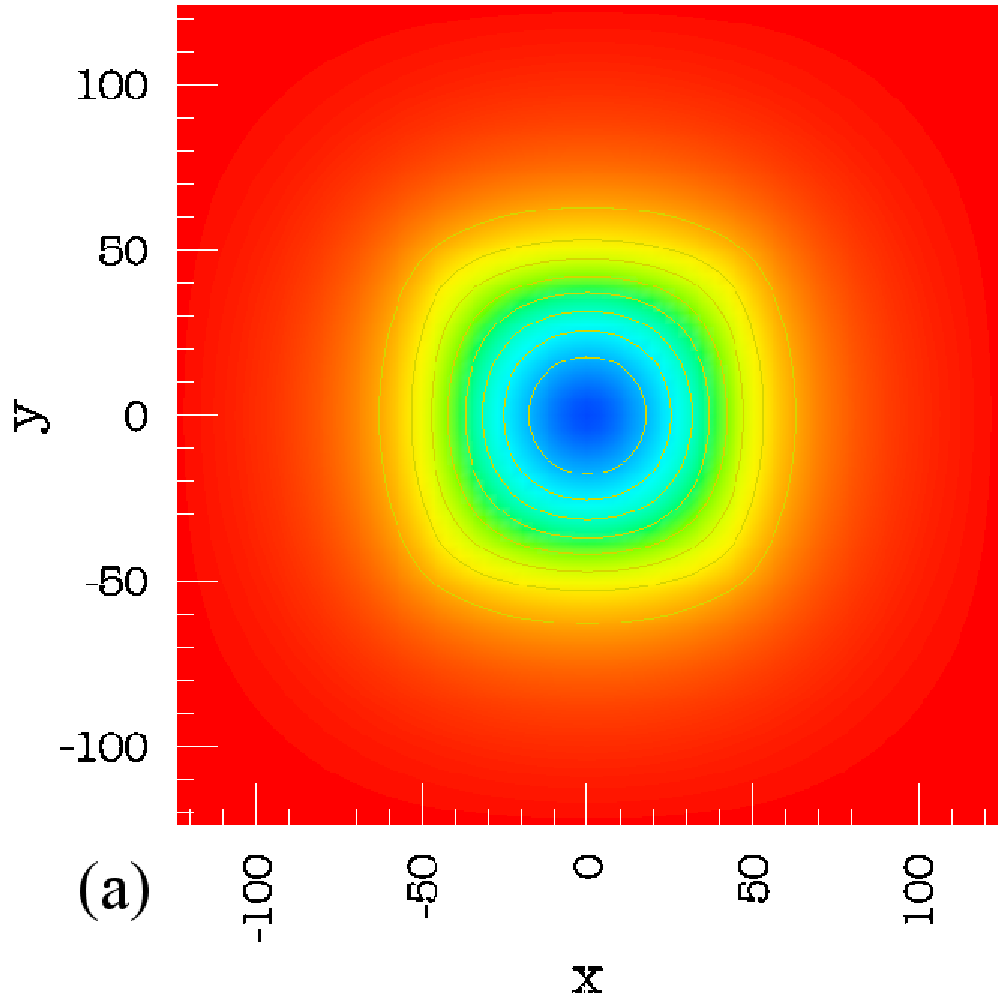}
\includegraphics[width=1.6in]{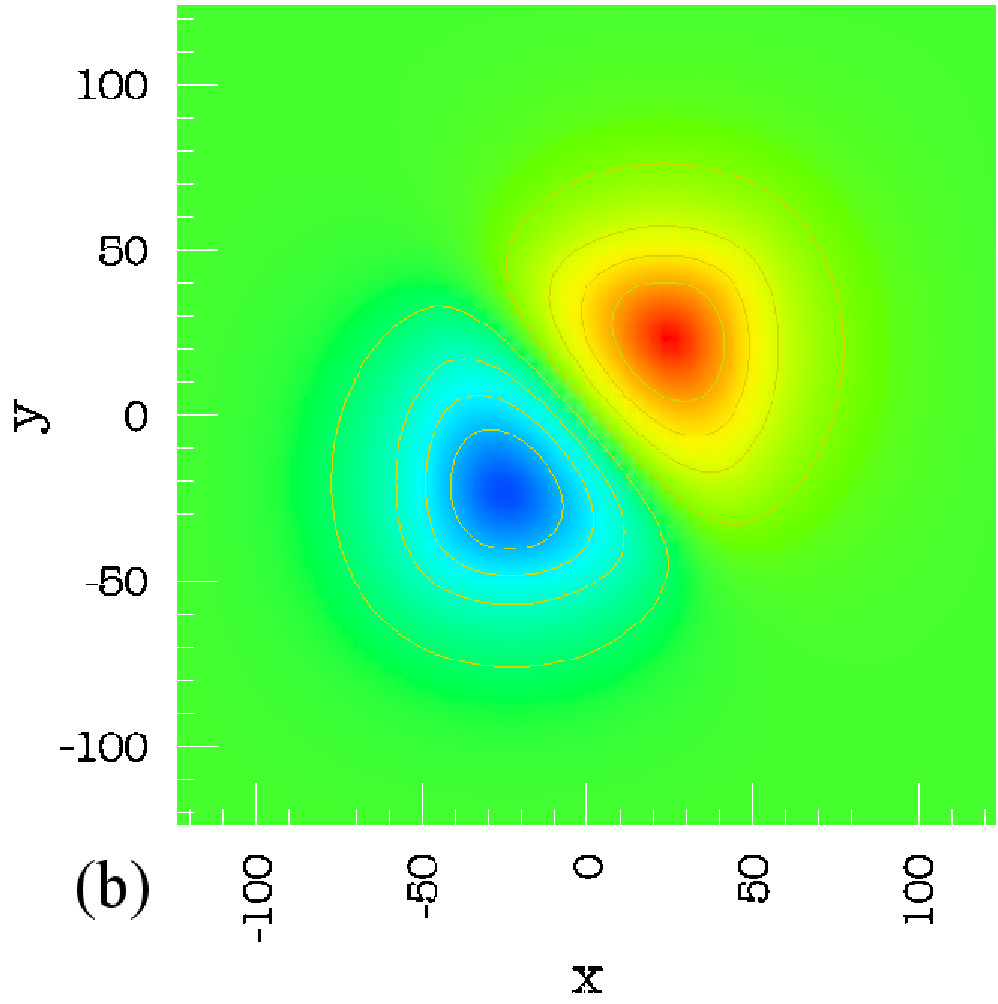}
\caption{\label{fig:trunc-wfs}Cross sections of (a) $\Psi_{\lambda_1}$ and (b) $\Psi_{\lambda_2}$ in the truncated-pyramid potential at $\text{z} = -15.5$ .}
\end{center}
\end{figure}
\begin{equation}
\label{eq:square-membrane}
\Psi_{(n,m)} \sim \sin (2n\pi x/a)\sin (2m\pi y/a).
\end{equation} 
Matching normal modes of wavefunctions in the pyramid and the truncated-pyramid with ones in the square membrane is straightforward only for the ground state. For their excited-states, it appears that the nodal line of $\lambda_2$ in the pyramid is along vertical with a bit distortion, while one in the truncated-pyramid is along diagonal. These patterns are due to the hybridization or superposition of the fundamental normal modes $(1,2)$ and $(2,1)$ of the square membrane. By configuring out a combination of $\Psi_{(1,2)}$ and $\Psi_{(2,1)}$, it yields the following equivalence in normal modes of the wavefunctions:\\
 For the pyramid, 
\begin{eqnarray}
&& \Psi_{\lambda_1}\sim \Psi_{(1,1)},~~\Psi_{\lambda_2} \sim -\Psi_{(1,2)}+4\Psi_{(2,1)},
 \nonumber   \\
&&\Psi_{\lambda_3} \sim 4\Psi_{(1,2)}+\Psi_{(2,1)}.
\end{eqnarray}
For the truncated-pyramid,
\begin{eqnarray}
&&\Psi_{\lambda_1}  \sim  \Psi_{(1,1)},~~\Psi_{\lambda_2} \sim -\Psi_{(1,2)}-\Psi_{(2,1)},
 \nonumber\\
&&\Psi_{\lambda_3} \sim  \Psi_{(1,2)}-\Psi_{(2,1)}. 
\end{eqnarray}
\end{itemize}
\end{enumerate}

\section{Conclusions}

Octopus is used to simulate the energy levels and wavefunctions in the geometrical potential wells. Although one can guess from the symmetry of the wells what the possible solutions including their degeneracies would look like, octopus is able to verify these presumptions. Since octopus is based on a real-space computation, it needs to take a close look at the effect of the spacing parameter to the energy eigenvalue before doing simulation. For the cone, pyramid and truncated-pyramid, there seems to have many spacing parameters producing approximately the same eigenvalue. So, spacing of 0.3875 nm is chosen as in Refs. \onlinecite{Hwang} and \onlinecite{Wang}. To simulate the electron energy levels in quantum mechanical scheme like the ones in parabolic band approximation scheme, the input files are run initially to find the electron mass fraction whose suitable value is 0.036. It yields the energy eigenvalues in the cone off from those in Ref. \onlinecite{Voss} in an order about 0.02--0.05 and ones in both the pyramid and the truncated-pyramid off from those in in Refs. \onlinecite{Hwang} and \onlinecite{Wang} in an order about 0.01. Octopus is also able to produce the DX-formatted wavefunctions. When looking at their cross sections, normal modes in the cone match exactly one--to--one with those in the circular membrane. In the pyramid and the truncated-pyramid, only normal mode in the ground state matches with that in the square membrane, while ones in their first excited-states are from the hybridization or superposition of fundamental modes $\Psi_{(1,2)}$ and $\Psi_{(2,1)}$ in the square membrane. Without the simulation, these vibrational normal modes might be beyond imagination.     

\begin{acknowledgments}
The author would like to thank Dr. Chittanon Buranachai for proofreading, discussion and criticism and all contributors of octopus, OpenDX, Xubuntu and etc. for devotion in creating and developing opensource software packages. This work was supported by the Department of Physics, Prince of Songkla University.
\end{acknowledgments}

\end{document}